\documentclass[3p,times,twocolumn,fleqn]{elsarticle}

\usepackage{ecrc}

\volume{00}

\firstpage{1}

\journalname{Nuclear Physics B Proceedings Supplements}

\runauth{O. Yasuda}

\jnltitlelogo{Nuclear Physics B Proceedings Supplements}

\usepackage{amssymb}

\usepackage{graphicx}

\begin{document}

\begin{frontmatter}


\dochead{}

\title{The KTY formalism and nonadiabatic contributions to the neutrino oscillation probability}


\author{Osamu Yasuda}

\address{Department of Physics, Tokyo Metropolitan University,
Minami-Osawa, Hachioji, Tokyo 192-0397, Japan}

\begin{abstract}

It is shown how to obtain the analytical expression
for the effective mixing angle in matter using the formalism which was
developed by Kimura, Takamura and Yokomakura.  If the baseline of the
neutrino path is long enough so that averaging over rapid oscillations
is a good approximation, then with the help of Landau's method,
the nonadiabatic contribution to the oscillation probability can be
expressed analytically by this formalism.  We give two examples in which
the present method becomes useful.
\end{abstract}


\end{frontmatter}



\section{Introduction}

Neutrino oscillation is a quantum mechanical
interference effect which sometimes has complex behaviors,
particularly in matter.  To discuss the behaviors of neutrino oscillation
intuitively, it is important to have analytical formulae for the
oscillation probability.  Unfortunately, an
analytical formula in the three flavor mixing scheme in matter
is quite complicated.  In
2002 Kimura, Takamura and Yokomakura (KTY) discovered a compact
formula~\cite{Kimura:2002hb,Kimura:2002wd} for the neutrino
oscillation probability in matter with constant density.  Subsequently
the KTY framework was
generalized to more general cases.  Ref.\,\cite{Zhang:2006yq} discussed
the four neutrino mixing scheme in matter with constant density.
Ref.\,\cite{FernandezMartinez:2007ms} discussed the case
with unitarity violation.
Ref.\,\cite{Yasuda:2007jp} discussed two cases of neutrino oscillation
in the adiabatic approximation,
the one with non-standard interactions where
the matter potential has non-diagonal elements in the flavor basis, or
the other with large neutrino magnetic moments in a magnetic field.

In general, however, adiabatic approximation may not be good, and in
this talk I discuss nonadiabatic contributions to the
oscillation probability.
When there are more than two neutrino mass eigenstates, there can be
more than one level crossing.
It is believed\,\footnote{See, e.g., Ref.\,\cite{Kuo:1989qe}
and references therein. See also Ref.\,\cite{Yamamoto:2010jc}
for a discussion on the condition to justify such a
treatment.}
that the nonadiabatic
contributions to the transition phenomena in a problem with three or
more eigenstates can be treated approximately well by applying the
method for two state problems
\cite{landau,Zener:1932ws} at each level crossing, if the
the two resonances are sufficiently far apart.
Throughout this talk I discuss the case in which
the baseline of the neutrino path is long enough
so that averaging over rapid oscillations
is a good approximation,
as in the case of the solar neutrino deficit phenomena.

\section{The
oscillation probability\label{
oscillation}}

\subsection{The oscillation probability in the adiabatic
approximation\label{adiabatic}}

The equation of motion for neutrinos propagating in matter with
general potential is given by
\begin{eqnarray}
i{d\Psi \over dt}=
\left[U{\cal E}_0U^{-1}
+{\cal A}(t)
\right]\Psi,
\label{sch1}
\end{eqnarray}
where
\begin{eqnarray}
{\cal E}_0&\equiv&{\mbox{\rm diag}}\left(E_1,E_2,E_3\right),
\nonumber\\
{\cal A}(t)&\equiv&
\left(
\begin{array}{ccc}
 A_{ee}(t) & A_{e\mu}(t) & A_{e\tau}(t)\\
 A_{\mu e}(t) & A_{\mu\mu}(t) & A_{\mu\tau}(t)\\
 A_{\tau e}(t) & A_{\tau\mu}(t) & A_{\tau\tau}(t)
\end{array}
\right).
\nonumber
\end{eqnarray}
Since the matrix which is proportional to identity gives
contribution only to the phase of the probability amplitude,
instead of ${\cal E}_0$ itself, we use the following quantity:
\begin{eqnarray}
{\cal E}\equiv{\cal E}_0-E_1\mbox{\bf 1}=\mbox{\rm diag}(0,\Delta E_{21},\Delta
E_{31}),
\nonumber
\end{eqnarray}
where 
$\Delta E_{jk}\equiv E_j - E_k \simeq (m_j^2-m_k^2)/2
|{\vec p}|.$
The $3\times3$ matrix on the
right hand side of Eq.\,(\ref{sch1}) can be formally
diagonalized as:
\begin{eqnarray}
U{\cal E}U^{-1}+{\cal A}(t)
=\tilde{U}(t)\tilde{{\cal E}}(t)\tilde{U}^{-1}(t),
\label{sch3}
\end{eqnarray}
where
$\tilde{{\cal E}}(t)\equiv{\mbox{\rm diag}}\left(
\tilde{E}_1(t),\tilde{E}_2(t),\tilde{E}_3(t)\right)$
is a diagonal matrix with the energy eigenvalues
$\tilde{E}_j(t)$ in the presence of the matter effect.
Substituting the diagonalized form (\ref{sch3})
of the Hamiltonian into the Dirac equation (\ref{sch1}),
we have
\mathindent=5mm
\begin{eqnarray}
i\frac{d\tilde{\Psi}}{dt}=
\left[\tilde{\cal E}-i\tilde{U}^{-1}
\left(\frac{d{\ }}{dt}\tilde{U}\right)
\right]\tilde{\Psi},
\label{sch4}
\end{eqnarray}
where $\tilde{\Psi}$ is the effective energy
eigenstate defined by
\begin{eqnarray}
\tilde{\Psi}(t)\equiv 
\left(
\begin{array}{c}
\tilde{\nu}_1(t)\\
\tilde{\nu}_2(t)\\
\tilde{\nu}_3(t)
\end{array}
\right)\equiv 
\tilde{U}^{-1}(t)\Psi(t).
\nonumber
\end{eqnarray}

If the term $\tilde{U}^{-1}d\tilde{U}/dt$ in
(\ref{sch4}) is negligible compared with
$\tilde{{\cal E}}$, i.e., if
adiabatic approximation is good, then
the oscillation probability is given by
\mathindent=-5mm
\begin{eqnarray}
&&
P(\nu_\alpha\rightarrow\nu_\beta)
=\sum_{j,k}\,\tilde{U}_{\beta j}(L)\tilde{U}^\ast_{\beta k}(L)
\tilde{U}^\ast_{\alpha j}(0)\tilde{U}_{\alpha k}(0)
\nonumber\\
&&\qquad\qquad\quad\times
\exp\left[-i\int_0^L\Delta\tilde{E}_{jk}(t)\,dt\right],
\label{proba1}
\end{eqnarray}
where we have defined
$\Delta \tilde{E}_{jk}(t)\equiv\tilde{E}_j(t)-\tilde{E}_k(t).$
The bilinear quantity
$\tilde{X}^{\alpha\beta}_j(t)\equiv
\tilde{U}_{\alpha j}(t)\tilde{U}^\ast_{\beta j}(t)$
can be expressed analytically
as\,\cite{Kimura:2002hb,Kimura:2002wd,Xing:2005gk,Yasuda:2007jp}
\mathindent=-7mm
\begin{eqnarray}
&&\left(\begin{array}{c}
\tilde{X}^{\alpha\beta}_1(t)\cr\cr
\tilde{X}^{\alpha\beta}_2(t)\cr\cr
\tilde{X}^{\alpha\beta}_3(t)
\end{array}\right)
=\left(\begin{array}{ccc}
\displaystyle
\frac{{\ }1}{\Delta \tilde{E}_{21} \Delta \tilde{E}_{31}}
(\tilde{E}_2\tilde{E}_3, & -(\tilde{E}_2+\tilde{E}_3),&
1)\cr
\displaystyle
\frac{-1}{\Delta \tilde{E}_{21} \Delta \tilde{E}_{32}}
(\tilde{E}_3\tilde{E}_1, & -(\tilde{E}_3+\tilde{E}_1),&
1)\cr
\displaystyle
\frac{{\ }1}{\Delta \tilde{E}_{31} \Delta \tilde{E}_{32}}
(\tilde{E}_1\tilde{E}_2, & -(\tilde{E}_1+\tilde{E}_2),&
1)\cr
\end{array}\right)
\nonumber\\
&&\qquad\qquad\quad\times\left(\begin{array}{r}
\delta_{\alpha\beta}\cr\cr
\left[U{\cal E}U^{-1}+{\cal A}(t)\right]_{\alpha\beta}\cr\cr
\left[\left(U{\cal E}U^{-1}+{\cal A}(t)\right)^2\right]_{\alpha\beta}
\end{array}\right),
\label{solx}
\end{eqnarray}
where the $t-$dependence of
the quantities $\tilde{E}_j$, $\Delta \tilde{E}_{jk}$
is suppressed for simplicity in Eq.\,(\ref{solx}).

\subsection{The nonadiabatic correction to the
oscillation probability\label{nonadiabatic}}

If the adiabatic approximation is not good,
on the other hand, then
Eq. (\ref{proba1}) should be modified by taking
non-adiabatic contributions into account.

\begin{figure}[tb]
\begin{center}
\vspace{-0.7cm}
\includegraphics[width=8.5cm]{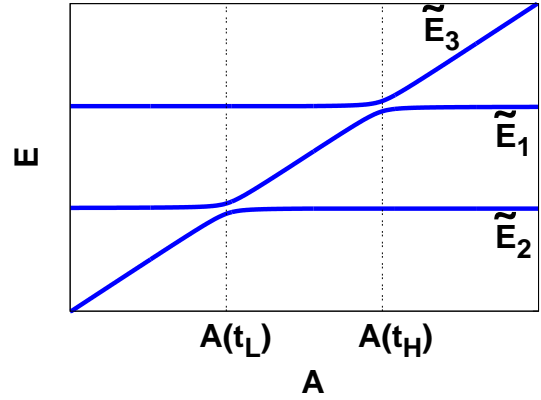}
\vspace{-0.cm}
\caption{\label{fig0} 
The three energy eigenvalues
$\tilde{E}_j~(j=1,2,3)$.  The potential
$A_{\alpha\beta}(t)$ is assumed to be
of the form diag($A(t)$,0,0) and
$A(t)$ is assumed to be linear in $t$
in this figure.  We have $\tilde{E}_3\simeq \tilde{E}_1$ near $t=t_H$
and $\tilde{E}_1\simeq \tilde{E}_2$ near $t=t_L$.}
\end{center}
\end{figure}

In the three flavor case, there are
at most two level crossings for neutrinos as in the case
of a supernova\,\cite{Dighe:1999bi}.  Let us assume
that there are two level crossings.
Let $t=t_H$ and $t=t_L$ be the level crossing points,
and we assume that the energy levels
1 (1) and 2 (3) cross at $t=t_L$ ($t=t_H$),
respectively (See Fig.~\ref{fig0}).

Except the two regions near $t=t_H$ and $t=t_L$,
we can integrate (\ref{sch4}) in the
adiabatic approximation.  Near the
two level crossings $t=t_H$ and $t=t_L$,
the energy eigenstates at $t=t_H\pm\epsilon$
and $t=t_L\pm\epsilon$
are related by the matrices $W_H$ and $W_L$
as in the case of the two level problem~\cite{Parke:1986jy}:
\mathindent=5mm
\begin{eqnarray}
&&\tilde{\Psi}(t_H+\epsilon)=W_H\tilde{\Psi}(t_H-\epsilon)
\nonumber\\
&&\tilde{\Psi}(t_L+\epsilon)=W_L\tilde{\Psi}(t_L-\epsilon)
\nonumber
\end{eqnarray}
In this case the flavor eigenstate is given by
\mathindent=-5mm
\begin{eqnarray}
&&\quad\Psi(L)
\nonumber\\
&&=\tilde{U}(L)
\exp\left[-i\int_{t_L}^L\,\tilde{{\cal E}}(t)\,dt\right]
\,W_L\,
\exp\left[-i\int_{t_H}^{t_L}\tilde{{\cal E}}(t)\,dt\right]
\,\nonumber\\
&&~\times
W_H\,\exp\left[-i\int_0^{t_H}\tilde{{\cal E}}(t)\,dt\right]
\tilde{U}(0)^{-1}\Psi(0).
\nonumber
\end{eqnarray}
Taking average over rapid oscillations,
we have the probability
\mathindent=-8mm
\begin{eqnarray}
&&\qquad P(\nu_\alpha\to\nu_\beta)
\nonumber\\
&&=\sum_{j,k,\ell}\,
\left|\tilde{U}(L)_{\beta j}\right|^2
\left|(W_L)_{jk}\right|^2
\left|(W_H)_{k\ell}\right|^2
\left|\tilde{U}(0)_{\alpha \ell}\right|^2
\nonumber\\
&&=\left(\begin{array}{ccc}
\left|U_{\beta 1}\right|^2&
\left|U_{\beta 2}\right|^2&
\left|U_{\beta 3}\right|^2
\end{array}\right)
\left(\begin{array}{ccc}
1-P_L & P_L& 0 \\
 P_L  & 1-P_L & 0\\
0 &0 & 1
\end{array}\right)\nonumber\\
&{\ }&\times
\left(\begin{array}{ccc}
1-P_H & 0 & P_H\\
0 & 1 & 0\\
P_H & 0 & 1-P_H
\end{array}\right)
\left(\begin{array}{c}
\left|\tilde{U}_{\alpha 1}(0)\right|^2\\
\left|\tilde{U}_{\alpha 2}(0)\right|^2\\
\left|\tilde{U}_{\alpha 3}(0)\right|^2
\end{array}\right).
\label{p3}
\end{eqnarray}
In (\ref{p3})
it was assumed that the nonadiabatic
contributions to the transition phenomena in a problem with three
eigenstates can be treated approximately well by applying the
method for two state problems at each level-crossing, when the
the two resonances are sufficiently far apart.
Using the WKB approximation\,\cite{landau},
the jumping factors in (\ref{p3}) are given by
\mathindent=0mm
\begin{eqnarray}
&&P_H= \exp\left[ -\,\mbox{\rm Im}\int_C\,
\Delta \tilde{E}_{31}(t)\,dt
\right],
\label{ph0}\\
&&P_L= \exp\left[ -\,\mbox{\rm Im}\int_C\,
\Delta \tilde{E}_{21}(t)\,dt
\right].
\label{pl0}
\end{eqnarray}

\subsection{The effective mixing angles in matter}
To evaluate the jumping factors $P_H$ and $P_L$
in the previous subsection, we assume here
as in the two flavor case that
the exponents in (\ref{ph0}) and (\ref{pl0})
are related by the ratio of
the difference of the energy eigenvalues of the two levels to
the derivative of the effective mixing angle
at the level-crossing.
For that purpose, it is necessary to
know the effective mixing angles at the level crossings.
In this subsection I show how to derive
the expression for the effective mixing angle in the presence
of the matter using the KTY formalism.
Our strategy here is to start with effective matrix elements
$\tilde{X}^{\alpha\beta}_j$ which are obtained by the KTY formalism
and to determine the phase of each element by demanding
that it be consistent with the standard parametrization
of the mixing matrix element:
\mathindent=5mm
\begin{eqnarray}
\tilde{U}=e^{i\tilde{\theta}_{23}\lambda_7}\,
\Gamma_{\tilde{\delta}}^{(13)}\,
e^{i\tilde{\theta}_{13}\lambda_5}\,
(\Gamma_{\tilde{\delta}}^{(13)})^{-1}\,
e^{i\tilde{\theta}_{12}\lambda_2},
\label{param1}
\end{eqnarray}
where $\lambda_j~(j=2,5,7)$ are
the Gell-Mann matrices defined by
$\lambda_2\equiv
\left(\begin{array}{ccc}
0&-i&0\cr
i&0&0\cr
0&0&0
\end{array}\right)$,
$\lambda_5\equiv
\left(\begin{array}{ccc}
0&0&-i\cr
0&0&0\cr
i&0&0
\end{array}\right)$,
$\lambda_7\equiv
\left(\begin{array}{ccc}
0&0&0\cr
0&0&-i\cr
0&i&0
\end{array}\right)$,
and
$\Gamma_\delta^{(13)}\equiv
\mbox{\rm diag}(e^{-i\delta/2},1,e^{i\delta/2})$.
From the identity
$\tilde{U}_{\alpha j}=e^{i\arg{\tilde{U}_{ej}}}\tilde{X}^{\alpha e}_j
/\sqrt{\tilde{X}^{ee}_j}$,
we can postulate the
form for $\tilde{U}$
\mathindent=5mm
\begin{eqnarray}
\tilde{U}\equiv e^{i\varphi_0}\,e^{i\varphi_3\lambda_3}\,
e^{i\varphi_9\lambda_9}\,\tilde{U}_0\,
e^{i\varphi'_9\lambda_9}\,e^{i\varphi'_3\lambda_3},
\label{utilde1}
\end{eqnarray}
where
$(\tilde{U}_0)_{\alpha j}\equiv
\tilde{X}^{\alpha e}_j/\sqrt{\tilde{X}^{ee}_j}$,
$\lambda_3\equiv\mbox{\rm diag}(1,-1,0)$,
$\lambda_9\equiv\mbox{\rm diag}(1,0,-1)$.
and choose the phases
$\varphi_0$, $\varphi_3$, $\varphi_9$,
$\varphi'_3$ and $\varphi'_9$ so that
the form (\ref{utilde1}) is consistent with
(\ref{param1}) (See Ref.~\cite{Yasuda:2014hwa}
for details).
Comparing Eqs.\,(\ref{param1}) and (\ref{utilde1}),
we find
\mathindent=-5mm
\begin{eqnarray}
&&\cos2\tilde{\theta}_{12}=\frac
{\tilde{X}^{ee}_1-\tilde{X}^{ee}_2}
{\tilde{X}^{ee}_1+\tilde{X}^{ee}_2},
\label{theta12tilde}\\
&&\cos2\tilde{\theta}_{13}=
1-2\tilde{X}^{ee}_3,
\label{theta13tilde}\\
&&\cos2\tilde{\theta}_{23}=\frac
{|\tilde{X}^{\tau e}_3|^2-|\tilde{X}^{\mu e}_3|^2}
{|\tilde{X}^{\tau e}_3|^2+|\tilde{X}^{\mu e}_3|^2},
\label{theta23tilde}\\
&&\tilde{\delta}
=-\arg\det\tilde{U}_0
+\arg\tilde{X}^{\mu e}_1
+\arg\tilde{X}^{\mu e}_3
+\arg\tilde{X}^{\tau e}_3.
\label{deltatilde}
\end{eqnarray}
The quantities $\tilde{X}^{\alpha\beta}_j$
and $\det\tilde{U}_0$ in
Eqs.\,(\ref{theta12tilde})--(\ref{deltatilde})
are expressed in closed form by the known variables
as is seen in Eq.\,(\ref{solx}) on the assumption
that analytical expressions for all the eigenvalues are known.\footnote{
Here we note that the effective mixing angles were given in the standard three flavor
case in Ref.\,\cite{Zaglauer:1988gz}, whereas
Eqs.\,(\ref{theta12tilde})--(\ref{deltatilde}) are the results
for the case of the general potential.}.

The standard parametrization (\ref{param1}) is not the only
one for $3\times 3$ unitary matrices,
and other parametrizations are possible
as is described in Appendix B of Ref.~\cite{Yasuda:2014hwa}.
In the three flavor case, there can be at most two 
level crossings.  Depending on which pair
of the energy eigenvalues gets close at each
level crossing, the relevant
effective mixing angle varies.
The appropriate parametrization is
the one in which the orthogonal
matrix, which mixes the two energy
eigenstates, is located on the most right-hand side
of the unitary matrix $U$, because
in such a parametrization
the diagonalized matrix looks like
$\cdots O(\tilde{\theta}_{jk})
\mbox{\rm diag}(\cdots,\tilde{E}_j,\cdots,\tilde{E}_k,\cdots)
O(\tilde{\theta}_{jk})^T\cdots$,
and it becomes clear that $\tilde{\theta}_{jk}$
in the orthogonal matrix $O(\tilde{\theta}_{jk})$
plays a role of the effective mixing angle
which mixes the energy eigenstates with
the energy $\tilde{E}_j$ and $\tilde{E}_k$.
Furthermore in order for the effective mixing angle
$\tilde{\theta}_{jk}$
to be consistent with the two flavor
description, $\tilde{\theta}_{jk}$ should
become maximal at the level crossing.
Thus, while the effective mixing
angle at $t=t_L$ is $\tilde{\theta}_{12}$
defined by (\ref{theta12tilde}) for the
standard parametrization (\ref{param1}),
the one at $t=t_H$
should be $\tilde{\varphi}_{13}$ defined by
\mathindent=5mm
\begin{eqnarray}
&&\cos2\tilde{\varphi}_{13}=\frac
{\tilde{X}^{ee}_1-\tilde{X}^{ee}_3}
{\tilde{X}^{ee}_1+\tilde{X}^{ee}_3},
\label{varphi12tilde}
\end{eqnarray}
for the parametrization
\begin{eqnarray}
&&
\tilde{U}=
e^{i\tilde{\varphi}_{23}\lambda_7}
\,\Gamma_\delta^{(12)}\,
e^{i\tilde{\varphi}_{12}\lambda_2}
\,(\Gamma_\delta^{(12)})^{-1}\,
e^{i\tilde{\varphi}_{13}\lambda_5},
\nonumber
\end{eqnarray}
where
$\Gamma_\delta^{(12)}\equiv
\mbox{\rm diag}(e^{-i\delta/2},e^{i\delta/2},1)$.
From the analogy with the two flavor case,
the jumping factors $P_H$ and $P_L$ are
given by
\mathindent=5mm
\begin{eqnarray}
&& P_H=
\displaystyle\exp\left(-\frac{\pi}{2}\,F\,
\frac{\Delta \tilde{E}_{31}}
{2\left|
d\tilde{\varphi}_{13}/dt
\right|_{t=t_H}}\right),
\label{ph}\\
&& P_L=
\displaystyle\exp\left(-\frac{\pi}{2}\,F\,
\frac{\Delta \tilde{E}_{21}}
{2\left|
d\tilde{\theta}_{12}/dt
\right|_{t=t_L}}\right).
\nonumber
\end{eqnarray}
$\tilde{\varphi}_{13}$ and
$\tilde{\theta}_{12}$ are
defined by (\ref{varphi12tilde}) 
and (\ref{theta12tilde}), and
$F$ is the factor which depends on the
form of the potential $A_{\alpha\beta}(t)$,
and $F=1$ in the case of
a linear potential.

\section{Examples}

Now I discuss two examples to demonstrate how the general discussions
in the previous section are applied.

  \subsection{The case with non-standard interactions
    \label{nonstandard}}

The first example is the oscillation probability
in the presence of new physics in 
propagation~\cite{Wolfenstein:1977ue,Guzzo:1991hi,Roulet:1991sm}.
In this case the mass matrix is given by
\begin{eqnarray}
U{\cal E}U^{-1}
+{\cal A}_{NP}
\label{matrixnp}
\end{eqnarray}
where
${\cal A}_{NP}\equiv\sqrt{2}G_FN_e
\left(
\begin{array}{ccc}
 1+\epsilon_{ee} & \epsilon_{e\mu} & \epsilon_{e\tau}\\
 \epsilon_{e\mu}^\ast & \epsilon_{\mu\mu} & \epsilon_{\mu\tau}\\
 \epsilon_{e\tau}^\ast & \epsilon_{\mu\tau}^\ast & \epsilon_{\tau\tau}
\end{array}
\right)$.
The dimensionless quantities
$\epsilon_{\alpha\beta}$ stand for possible deviation from
the standard matter effect.
It is known~\cite{Davidson:2003ha}
that the constraints on the
parameters $\epsilon_{e\mu}$, $\epsilon_{\mu\mu}$,
$\epsilon_{\mu\tau}$ are strong
($|\epsilon_{\alpha\mu}|\simeq {\cal O}(10^{-2})~(\alpha=e,\mu,\tau$)
while those on the parameters
$\epsilon_{ee}, \epsilon_{e\tau}, \epsilon_{\tau\tau}$
are weak
($|\epsilon_{ee}|,~|\epsilon_{e\tau}|, |\epsilon_{\tau\tau}|
\simeq {\cal O}(1)$.
In Ref.\,\cite{Friedland:2005vy} it was found that large values
($\sim {\cal O}(1)$) of the
parameters $\epsilon_{ee}, \epsilon_{e\tau}, \epsilon_{\tau\tau}$ are
consistent with all the experimental data including those of the
atmospheric neutrino data, provided that one of the
eigenvalues of the matrix (\ref{matrixnp}) at high energy limit
becomes zero, and that such a constraint implies
the relation $\epsilon_{\tau\tau}\simeq|\epsilon_{e\tau}|^2/(1+\epsilon_{ee})$.
For simplicity, therefore, we consider the potential matrix
\mathindent=0mm
\begin{eqnarray}
&&{\cal A}_{NP}=A
\left(
\begin{array}{ccc}
 1+\epsilon_{ee} & 0 & \epsilon_{e\tau}\\
 0 & 0 & 0\\
 \epsilon_{e\tau}^\ast & 0 & |\epsilon_{e\tau}|^2/(1+\epsilon_{ee})
\end{array}
\right).
\label{potentialnp}
\end{eqnarray}
Then ${\cal A}_{NP}$ can be diagonalized as
\mathindent=5mm
\begin{eqnarray}
{\cal A}_{NP} =　e^{i\gamma'\lambda_9}e^{-i\beta\lambda_5}\,
\mbox{\rm diag}\left(\lambda_{e'},0,0\right)
e^{i\beta\lambda_5}e^{-i\gamma'\lambda_9},
\label{np2}
\end{eqnarray}
where
$\tan\beta=|\epsilon_{e\tau}|/
(1+\epsilon_{ee})$,
$\gamma'\equiv\mbox{\rm arg}\,(\epsilon_{e\tau})/2$,
$\lambda_{e'}=
A(1+\epsilon_{ee})/\cos^2\beta$.
The mass matrix (\ref{matrixnp})
can be written as
\mathindent=-8mm
\begin{eqnarray}
&&
~~~U{\cal E}U^{-1}+{\cal A}_{NP}
\nonumber\\
&&=
e^{i\gamma'\lambda_9}e^{-i\beta\lambda_5}
e^{-i\phi_9\lambda_9}
e^{-i\phi_3\lambda_3}
\left[
U''{\cal E}U''^{-1}
+\mbox{\rm diag}\left(\lambda_{e'},0,0\right)
\right]
\nonumber\\
&&~~\times e^{-i\omega_3\lambda_3}
e^{-i\omega_9\lambda_9}
e^{i\beta\lambda_5}e^{-i\gamma'\lambda_9},
\label{matrixnp2}
\end{eqnarray}
where the phases $\phi_3$, $\phi_9$,
$\omega_3$ and $\omega_9$, which are
defined in Appendix C of Ref.~\cite{Yasuda:2014hwa},
are introduced to make $U''$
consistent with the standard parametrization
(\ref{param1}).  The expressions for th mixing angles
$\theta''_{jk}$ and the CP phase $\delta''$
in the standard parametrization of $U''$
are given in Ref.~\cite{Yasuda:2014hwa}.
Among others the mixing angles
$\theta''_{12}$ and $\theta''_{13}$ are
given by
\mathindent=0mm
\begin{eqnarray}
&&\theta''_{12}=
\tan^{-1}\frac
{|c_\beta e^{-i\gamma'}U_{e2}+s_\beta e^{i\gamma'}U_{\tau2}|}
{|c_\beta e^{-i\gamma'}U_{e1}+s_\beta e^{i\gamma'}U_{\tau1}|},
\nonumber\\
&&\theta''_{13}=
\sin^{-1}|c_\beta e^{-i\gamma'}U_{e3}+s_\beta e^{i\gamma'}U_{\tau3}|,
\nonumber
\end{eqnarray}
where $c_\beta\equiv\cos\beta$, $s_\beta\equiv\sin\beta$.
The inside of the square bracket in the
mass matrix (\ref{matrixnp2}) has exactly
the same form as that of the standard case
with replacement $\theta_{jk}\to\theta''_{jk}$,
$\delta\to\delta''$ and $A\to\lambda_{e'}$.
Furthermore, at the two level-crossings specified by
$\Delta E_{31}\cos2\theta''_{13} = \lambda_{e'}$ and
$\Delta E_{21}\cos2\theta''_{12} = (c''_{13})^2\lambda_{e'}$,
$\tilde{\theta}''_{13}$ and
$\tilde{\theta}''_{12}$ become $\pi/4$, respectively.
Therefore, $\tilde{\theta}''_{13}$ and $\tilde{\theta}''_{12}$ can
be regarded as the appropriate mixing angles
to describe the nonadiabatic transition
at the two level-crossings.
Hence we can deduce the jumping factors
at the two level-crossings\,\footnote{
The quantity $P_L$ was given first in
Ref.\,\cite{oai:arXiv.org:hep-ph/0402266}
whose result agrees with ours.}:
\mathindent=-2mm
\begin{eqnarray}
&&P_H=\exp\left(
-\frac{\pi}{2}\cdot\frac{\Delta E_{31}\sin^22\theta''_{13}}
{\cos2\theta''_{13}|d\log A/dt|_{\mbox{\rm\scriptsize resonance}}}
\right)
\label{phnp}\\
&&P_L=\exp\left(
-\frac{\pi}{2}\cdot\frac{\Delta E_{21}\sin^22\theta''_{12}}
{\cos2\theta''_{12}|d\log A/dt|_{\mbox{\rm\scriptsize resonance}}}
\right)
\label{plnp}
\end{eqnarray}

To estimate the effective mixing matrix elements at the origin $L=0$,
we assume that the matter effect $A$ is much larger than
the energy difference $|\Delta E_{jk}|$.
In this case we can ignore
the term ${\cal E}$ in Eq.\,(\ref{matrixb2}),
and Eq.\,(\ref{np2}) indicates that the mixing matrix
$\tilde{U}$ is given by $e^{i\gamma'\lambda_9}e^{-i\beta\lambda_5}$,
and we get
\begin{eqnarray}
&&|\tilde{U}_{\alpha j}(0)|^2=\left(
\begin{array}{ccc}
c_\beta^2&0&s_\beta^2\cr
0&1&0\cr
s_\beta^2&0&c_\beta^2
\end{array}\right).
\label{u0np}
\end{eqnarray}
From Eqs.\,(\ref{p3}), (\ref{phnp}), (\ref{plnp}) and (\ref{u0np}),
we can obtain the transition probability
$P(\nu_\alpha\to\nu_\beta)$ in the case with
the nonstandard neutrino interaction in propagation.

\subsection{The case with large magnetic moments and a magnetic field
  \label{magnetic}}

The second example is the case where
there are three active neutrinos with
magnetic moments and a large magnetic
field\,\footnote{
The possibility that the magnetic moments of neutrinos in
a large magnetic field affect the neutrino flavor transition
caught a lot of attention after this idea was applied to the
solar neutrino deficit in
Refs.\,\cite{Cisneros:1970nq,Okun:1986na,Lim:1987tk,Akhmedov:1988uk}.}.
This is an example where the energy eigenvalues
cannot be expressed as roots of a quadratic equation, and this
case demonstrates the usefulness of the KTY formalism.
Here we assume the magnetic interaction of Majorana type
$\mu_{\alpha\beta}\bar{\nu}_\alpha\,F_{\lambda\kappa}
\sigma^{\lambda\kappa}\,\nu^c_\beta + h. c.$,
and in this case the magnetic moments $\mu_{\alpha\beta}$
are real and anti-symmetric in flavor
indices: $\mu_{\alpha\beta}=-\mu_{\beta\alpha}$.
\mathindent=5mm
\begin{eqnarray}
{\cal M}\equiv
\left(
\begin{array}{cc}
U{\cal E}U^{-1}&{\cal B}\\
{\cal B}^\dagger&U^\ast{\cal E}(U^\ast)^{-1}
\end{array}\right)
\label{matrixb}
\end{eqnarray}
with
${\cal B}_{\alpha\beta}\equiv B\,\mu_{\alpha\beta}$
is the hermitian mass matrix
for neutrinos and anti-neutrinos without the matter effect
where neutrinos have the magnetic moments $\mu_{\alpha\beta}$
in the magnetic field $B$.

For simplicity we consider the limit
$\theta_{13}\to0$ and $\Delta m^2_{21}\rightarrow0$,
and we assume that all the CP phases vanish.
Then the matrix (\ref{matrixb}) can be rewritten as
\mathindent=-5mm
\begin{eqnarray}
&&{\cal M}
=\frac{1}{2}
\left(\begin{array}{rr}
{\bf 1}&i{\bf 1}\cr
i{\bf 1}&{\bf 1}
\end{array}\right)
\left(\begin{array}{cc}
U{\cal E}U^{-1}+i{\cal B}&0\cr
0&U{\cal E}U^{-1}-i{\cal B}
\end{array}\right)
\nonumber\\
&&\qquad\times\left(\begin{array}{rr}
{\bf 1}&-i{\bf 1}\cr
-i{\bf 1}&{\bf 1}
\end{array}\right),
\nonumber
\end{eqnarray}
so the problem of diagonalizing the $6\times6$ matrix (\ref{matrixb})
is reduced to diagonalizing the $3\times3$ matrices
$U{\cal E}U^{-1}\pm i{\cal B}$.
Since we are assuming that all the CP phases vanish,
all the matrix elements $U_{\alpha j}$ and
${\cal B}_{\alpha\beta}=-{\cal B}_{\beta\alpha}$
are real, $U{\cal E}U^{-1}\pm i{\cal B}$ can be diagonalized by
a unitary matrix and its complex conjugate:
\mathindent=5mm
\begin{eqnarray}
&&U{\cal E}U^{-1}+ i{\cal B}=
\tilde{U}\tilde{{\cal E}}\tilde{U}^{-1}
\label{matrixb2}\\
&&U{\cal E}U^{-1}- i{\cal B}=
\tilde{U}^\ast\tilde{{\cal E}}(\tilde{U}^\ast)^{-1},
\nonumber
\end{eqnarray}
and the equation for motion is given by
\mathindent=-5mm
\begin{eqnarray}
&&~~~ i\frac{d{\ }}{dt}
\left(\begin{array}{c}
\Psi(t)+i\Psi^c(t)\cr
\Psi(t)-i\Psi^c(t)
\end{array}\right)
\nonumber\\
&&=
\left(\begin{array}{c}
\tilde{U}(t)\tilde{{\cal E}}(t)\tilde{U}^{-1}(t)
\left\{\Psi(t)+i\Psi^c(t)\right\}
\cr
\tilde{U}^\ast(t)\tilde{{\cal E}}(t)(\tilde{U}^\ast)^{-1}(t)
\left\{\Psi(t)-i\Psi^c(t)\right\}
\end{array}\right).
\nonumber
\end{eqnarray}

\begin{figure}[tb]
\begin{center}
\vspace{-0.5cm}
\hspace{-0.3cm}
\includegraphics[width=9.0cm]{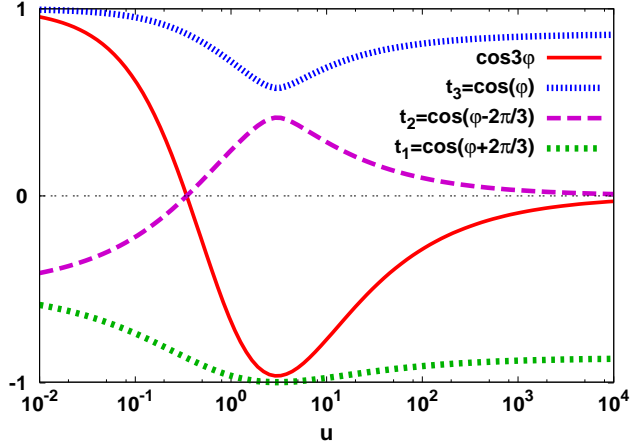}
\caption{\label{fig1} 
The behaviors of the normalized eigenvalues
$t_j\equiv\tilde{E}_j/\left(2
\sqrt{\Delta E_{31}^2/9+\Lambda^2/3}\,\right)
=\cos(\varphi+2j\pi/3)\,(j=1,2,3)$
and $\cos3\varphi$
as functions of
$u\equiv 3\Lambda^2/\Delta E_{31}^2$.
See Ref.~\cite{Yasuda:2014hwa} for details.}
\end{center}
\end{figure}

Introducing the notations
\begin{eqnarray}
&&{\cal B}_{\alpha\beta}=B\mu_{\alpha\beta}
\equiv\left(
\begin{array}{ccc}
0&-p_0&-q_0\cr
p_0&0&-r_0\cr
q_0&r_0&0
\end{array}\right),
\nonumber\\
\nonumber\\
&&~~~e^{-i{\theta}_{23}\lambda_7}\,{\cal B}\,
e^{i{\theta}_{23}\lambda_7}
\nonumber\\
&&=
\left(
\begin{array}{ccc}
0&-p_0c_{23}+q_0s_{23}&-p_0s_{23}-q_0c_{23}\cr
p_0c_{23}-q_0s_{23}&0&-r_0\cr
p_0s_{23}+q_0c_{23}&r_0&0
\end{array}\right)
\nonumber\\
&&\equiv\left(
\begin{array}{ccc}
0&-p&-q\cr
p&0&-r\cr
q&r&0
\end{array}\right),
\nonumber
\end{eqnarray}
it can be shown~\cite{Yasuda:2014hwa}
that Eq.\,(\ref{matrixb2}) can be rewritten as
\begin{eqnarray}
&&~~e^{i{\theta}_{23}\lambda_7}\,
e^{i\omega\lambda_2}\,
\left[\mbox{\rm diag}(0,0,\Delta E_{31})\,
+ \Lambda
e^{i\chi\lambda_5}\,
\lambda_2\,
e^{-i\chi\lambda_5}\,
\right]
\nonumber\\
&&\times e^{-i\omega\lambda_2}\,
e^{-i{\theta}_{23}\lambda_7}\,,
\label{matrixb4}
\end{eqnarray}
where $\Lambda$, $\omega$ and $\chi$ are
defined by
$\Lambda\equiv\sqrt{p^2+q^2+r^2}$,
$\omega\equiv\tan^{-1}(r/q)$,
$\chi\equiv\tan^{-1}(\sqrt{q^2+r^2}/p)$.
Since the we are mainly interested in the effective
mixing angle which mixes the two energy
eigenstates, the matrices $e^{i{\theta}_{23}\lambda_7}e^{i\omega\lambda_2}$
on the left-hand side and
$e^{-i\omega\lambda_2}e^{-i{\theta}_{23}\lambda_7}$
on the right-hand side of
the square bracket in Eq.\,(\ref{matrixb4}) are irrelevant, so
we discuss the following matrix:
\mathindent=-7mm
\begin{eqnarray}
&&{\cal M}\equiv\mbox{\rm diag}(0,0,\Delta E_{31})\,
-\frac{\Delta E_{31}}{3}\mbox{\bf 1}
+ \Lambda
e^{i\chi\lambda_5}\,
\lambda_2\,
e^{-i\chi\lambda_5},
\label{m0}
\end{eqnarray}
where a matrix which is proportional to identity
was subtracted for convenience in later calculations
so that the trace of ${\cal M}$ vanishes.
The eigenvalues $\tilde{E}_j$ of the matrix ${\cal M}$
are given by $\tilde{E}_j=
2\sqrt{\Delta E_{31}^2/9+\Lambda^2/3}\,
\cos(\varphi+2j\pi/3)$ $(j=1,2,3)$,
where
$\cos3\varphi\equiv
\left\{\left(\Delta E_{31}/3\right)^3
-(1+3\cos2\chi)\,\Lambda^2\Delta E_{31}/12
\right\}/
\left(\Delta E_{31}^2/9+\Lambda^2/3\right)^{3/2}$.
In Fig.~\ref{fig1}
the three eigenvalues $t_j~(j=1,2,3)$ which are normalized by
$2\sqrt{\Delta E_{31}^2/9+\Lambda^2/3}$ are depicted as a
function of $u\equiv 3\Lambda^2/\Delta E_{31}^2$.
If $\chi$ is small, then the two of the three
eigenvalues get close to each other, and
$\chi$ can be regarded as the vacuum mixing angle
near the level-crossing in the present case.
In this example, for a large value of $\Lambda\gg \Delta E_{31}$,
the energy eigenvalues are
$0$ and $\pm \Lambda$,
and we found that there is only one
level-crossing for
$|\Delta E_{31}|\sim \Lambda$,
unlike in the standard three flavor case.
So in the following we discuss the contribution
from one level-crossing only.

Furthermore,
it can be shown~\cite{Yasuda:2014hwa}
that the following relation holds:
\begin{eqnarray}
&&P(\nu_\alpha\rightarrow\nu_\beta)+P(\nu_\alpha\rightarrow\bar{\nu}_\beta)
\nonumber\\
&&=\sum_{j,k}
|\tilde{U}_{\beta j}(L)|^2\,
|(W_H)_{jk}|^2\,
|\tilde{U}_{\alpha k}^\ast(0)|^2
\nonumber\\
&&=\left(\begin{array}{ccc}
\left|U_{\beta 1}\right|^2&
\left|U_{\beta 2}\right|^2&
\left|U_{\beta 3}\right|^2
\end{array}\right)
\left(\begin{array}{ccc}
1 & 0 &0 \\
0 & 1-P_H & P_H\\
0 & P_H & 1-P_H
\end{array}\right)
\nonumber\\
&&\times\left(\begin{array}{c}
\left|\tilde{U}_{\alpha 1}(0)\right|^2\\
\left|\tilde{U}_{\alpha 2}(0)\right|^2\\
\left|\tilde{U}_{\alpha 3}(0)\right|^2
\end{array}\right),
\label{relationb}
\end{eqnarray}
where we have assumed that the
level-crossing occurs between
the energy eigenstates 2 and 3,
and we have assumed that
there is no magnetic field at the
endpoint $t=L$, and $|\tilde{U}_{\alpha j}(0)|^2$ in the
transition probability can be obtained
from the KTY formula (\ref{solx}).

In the approximation of the small mixing angle $\chi$,
the jumping factor $P_H$ can be calculated as~\cite{Yasuda:2014hwa}
\mathindent=5mm
\begin{eqnarray}
P_H&\simeq& \exp\left( -\frac{\pi}{|d\Lambda/dt|_{u=u_0}}\,\Delta E_{31}^2\,\chi^2
\right).
\label{phb}
\end{eqnarray}
It can be also shown that
the exponent of the
jumping factor $P_H$ coincides with
$-\pi/2$ times the $\gamma$ factor
in the case of a linear potential ($F=1$):
\begin{eqnarray}
\gamma = \left.\frac{\Delta \tilde{E}_{32}}
{2|d\tilde{\psi}_{23}/dt|}\right|_{u=u_0}
\simeq\frac{2\Delta E_{31}^2\chi^2}
{|d\Lambda/dt|_{u=u_0}}
=-\frac{\log P_H}{\pi/2}.
\nonumber
\end{eqnarray}

\section{Conclusions \label{conclusions}}
Using the formalism which was developed by Kimura, Takamura and
Yokomakura to express analytically the combination
$\tilde{X}^{\alpha\beta}_j\equiv
\tilde{U}_{\alpha j}\tilde{U}_{\beta j}^\ast$ of
the mixing matrix elements in matter with constant density,
we have shown that the effective mixing angle
can be analytically expressed in terms of
the mixing matrix elements in vacuum and the energy eigenvalues.
The analytical expression for the effective
mixing angle enables us to evaluate the
nonadiabatic contribution to the transition
probability based on the two assumptions:
(i) The nonadiabatic transitions in the case
with more than two energy eigenstates can
be separately treated as a two state problem
at each level crossing.
(ii) The exponent of the probability obtained
by the WKB method is proportional to the factor
$\gamma$ which is the ratio of the energy
difference of the two eigenstates to the derivative
of the effective mixing angle at the level crossing.
We have given two examples: one with
flavor dependent nonstandard interactions
in neutrino propagation and the other with
magnetic moments in a large magnetic field.

\section*{Acknowledgments}
This research was partly supported by a Grant-in-Aid for Scientific
Research of the Ministry of Education, Science and Culture, under
Grants No. 24540281 and No. 25105009.





\begin{thebibliography}{00}

\bibitem{Kimura:2002hb}
K.~Kimura, A.~Takamura and H.~Yokomakura,
Phys.\ Lett.\ B {\bf 537}, 86 (2002)
[arXiv:hep-ph/0203099].

\bibitem{Kimura:2002wd}
K.~Kimura, A.~Takamura and H.~Yokomakura,
Phys.\ Rev.\ D {\bf 66}, 073005 (2002)
[arXiv:hep-ph/0205295].

\bibitem{Zhang:2006yq}
  H.~Zhang,
  Mod.\ Phys.\ Lett.\  A {\bf 22}, 1341 (2007)
  [arXiv:hep-ph/0606040].

\bibitem{FernandezMartinez:2007ms} 
  E.~Fernandez-Martinez, M.~B.~Gavela, J.~Lopez-Pavon and O.~Yasuda,
  Phys.\ Lett.\ B {\bf 649}, 427 (2007)
  [hep-ph/0703098].

\bibitem{Yasuda:2007jp}
  O.~Yasuda,
  arXiv:0704.1531 [hep-ph].

\bibitem{Kuo:1989qe}
  T.~K.~Kuo and J.~T.~Pantaleone,
  Rev.\ Mod.\ Phys.\  {\bf 61}, 937 (1989).

\bibitem{Yamamoto:2010jc} 
  K.~Yamamoto,
  arXiv:1003.2853 [hep-ph].

\bibitem{landau}
L. Landau,
Phys. Z. Sowj. 2, 46 (1932).

\bibitem{Zener:1932ws}
  C.~Zener,
  Proc.\ Roy.\ Soc.\ Lond.\  A {\bf 137}, 696 (1932).

\bibitem{Xing:2005gk}
  Z.~z.~Xing and H.~Zhang,
  Phys.\ Lett.\  B {\bf 618}, 131 (2005)
  [arXiv:hep-ph/0503118].

\bibitem{Dighe:1999bi} 
  A.~S.~Dighe and A.~Y.~Smirnov,
  Phys.\ Rev.\ D {\bf 62}, 033007 (2000)
  [hep-ph/9907423].

\bibitem{Parke:1986jy} 
  S.~J.~Parke,
  Phys.\ Rev.\ Lett.\  {\bf 57}, 1275 (1986).

\bibitem{Yasuda:2014hwa}
  O.~Yasuda,
  Phys.\ Rev.\ D {\bf 89} (2014) 093023
  [arXiv:1402.5569 [hep-ph]].

\bibitem{Zaglauer:1988gz}
  H.~W.~Zaglauer and K.~H.~Schwarzer,
  Z.\ Phys.\  C {\bf 40}, 273 (1988).

\bibitem{Wolfenstein:1977ue} 
  L.~Wolfenstein,
  Phys.\ Rev.\ D {\bf 17}, 2369 (1978).

\bibitem{Guzzo:1991hi}
M.~M.~Guzzo, A.~Masiero and S.~T.~Petcov,
Phys.\ Lett.\ B {\bf 260}, 154 (1991);
%
\bibitem{Roulet:1991sm}
E.~Roulet,
Phys.\ Rev.\ D {\bf 44}, 935 (1991).

\bibitem{Davidson:2003ha}
  S.~Davidson, C.~Pena-Garay, N.~Rius and A.~Santamaria,
  JHEP {\bf 0303}, 011 (2003)
  [arXiv:hep-ph/0302093].

\bibitem{Friedland:2005vy}
  A.~Friedland and C.~Lunardini,
  Phys.\ Rev.\  D {\bf 72}, 053009 (2005)
  [arXiv:hep-ph/0506143].

\bibitem{oai:arXiv.org:hep-ph/0402266} 
  A.~Friedland, C.~Lunardini and C.~Pena-Garay,
  Phys.\ Lett.\ B {\bf 594}, 347 (2004)
  [hep-ph/0402266].

\bibitem{Cisneros:1970nq} 
  A.~Cisneros,
  Astrophys.\ Space Sci.\  {\bf 10}, 87 (1971).

\bibitem{Okun:1986na} 
  L.~B.~Okun, M.~B.~Voloshin and M.~I.~Vysotsky,
  Sov.\ Phys.\ JETP {\bf 64}, 446 (1986)
  [Zh.\ Eksp.\ Teor.\ Fiz.\  {\bf 91}, 754 (1986)].

\bibitem{Lim:1987tk} 
  C.~-S.~Lim and W.~J.~Marciano,
  Phys.\ Rev.\ D {\bf 37}, 1368 (1988).

\bibitem{Akhmedov:1988uk} 
  E.~K.~Akhmedov,
  Phys.\ Lett.\ B {\bf 213}, 64 (1988).

\end{thebibliography}
\end{document}